\documentclass[prd,nofootinbib,preprint,superscriptaddress]{revtex4}
\pdfoutput=1
\usepackage[T1]{fontenc}
\usepackage{amsmath,amssymb}
\usepackage{epsfig}
\usepackage{subfigure}
\usepackage{float}
\usepackage{graphicx}
\usepackage[usenames,dvipsnames]{color}
\usepackage{slashed}
\usepackage{multirow}
\usepackage[colorlinks,citecolor=blue]{hyperref}
\usepackage{pdfpages}
\usepackage{color}
\usepackage{comment}

\def\lsim{\raise0.3ex\hbox{$\;<$\kern-0.75em\raise-1.1ex\hbox{$\sim\;$}}}
\def\gsim{\raise0.3ex\hbox{$\;>$\kern-0.75em\raise-1.1ex\hbox{$\sim\;$}}}

\newcommand{\be}{\begin{equation}}
\newcommand{\ee}{\end{equation}}
\newcommand{\bea}{\begin{eqnarray}}
\newcommand{\eea}{\end{eqnarray}}

\graphicspath{{Figures/}}
\begin{document}

\title{Long lived inert Higgs boson in a fast expanding universe and its imprint on the cosmic microwave background}

\author{Dilip Kumar Ghosh}
\email{tpdkg@iacs.res.in}
\affiliation{School of Physical Sciences, Indian Association for the Cultivation of Science,\\ 2A $\&$ 2B Raja S.C. Mullick Road, Kolkata 700032, India}

\author{Sk Jeesun}
\email{skjeesun48@gmail.com}
\affiliation{School of Physical Sciences, Indian Association for the Cultivation of Science,\\ 2A $\&$ 2B Raja S.C. Mullick Road, Kolkata 700032, India}

\author{Dibyendu Nanda}
\email{dnanda@kias.re.kr}
\affiliation{School of Physical Sciences, Indian Association for the Cultivation of Science,\\ 2A $\&$ 2B Raja S.C. Mullick Road, Kolkata 700032, India}
\affiliation{School of Physics, Korea Institute for Advanced Study, Seoul 02455, Korea}

\begin{abstract}
Presence of any extra radiation energy density at the time of cosmic 
microwave background formation can significantly impact the measurement 
of the effective relativistic neutrino degrees of freedom or 
${\rm \Delta N_{eff}}$ which is very precisely measured by the 
Planck collaboration. Here, we propose a scenario 
where a long lived inert scalar, which is very weakly coupled to dark sector, 
decays to a fermion dark matter via {\it freeze-in} mechanism 
plus standard model neutrinos at very low temperature $(T< T_{\rm BBN})$. 
We explore this model in the fast expanding universe, where it is
assumed that the early epoch $(T > T_{\rm BBN})$ 
of the Universe is dominated by a non-standard
species $\Phi$ instead of the standard radiation. In this non-standard
cosmological picture, such late time decay of the inert scalar 
can inject some entropy to the neutrino sector after it decouples 
from the thermal bath and this will make substantial contribution to  
${\rm \Delta{N_{eff}}}$. Besides, in this scenario, the new 
contribution to $\Delta N_{\rm eff}$ is highly correlated with the 
dark matter sector. Thus, one can explore such {\it 
feebly} interacting dark matter particles (FIMP) by the precise measurement 
of $\Delta N_{\rm eff}$ using the current (Planck2018)
and forthcoming (CMB-S4 \& SPT3G) experiments.

 \end{abstract}
\maketitle

\section{Introduction} 
\label{sec:intro}

The discovery of the 125 GeV Higgs boson by both ATLAS and CMS 
collaborations at the LHC completes the basic building blocks of 
standard model (SM) particle physics, albeit some theoretical and experimental
shortcomings. The SM with its present setup is unable to explain the 
observed non-zero neutrino masses and mixings\cite{T2K:2011ypd,
DoubleChooz:2011ymz,DayaBay:2012fng,RENO:2012mkc,MINOS:2013xrl,
ParticleDataGroup:2020ssz} and the existence of dark matter 
as indicated by various astrophysical and cosmological measurements 
\cite{Zwicky:1933gu,Rubin:1970zza,Clowe:2006eq,Planck:2018vyg}. 
The resolution of these two fundamental puzzles of particle and 
astroparticle physics beg for an extension of the SM and 
a plethora of beyond the standard 
model (BSM) scenarios have been proposed to address these two issues. 
The neutrino masses and their mixing angles can be easily 
accommodated at tree level in the three Seesaw
mechanisms \cite{Minkowski:1977sc,Mohapatra:1979ia,
Schechter:1980gr,Gell-Mann:1979vob, Mohapatra:1980yp,Lazarides:1980nt,
Wetterich:1981bx,Schechter:1981cv,Brahmachari:1997cq,Foot:1988aq}.
Note that besides the tree-level seesaw mechanisms, small
neutrino masses can also be generated radiatively at one loop
level \cite{Ma:1998dn,Ma:2006km}. 

For a very long time, weakly interacting massive neutral particles 
(WIMP) \cite{Kolb:1990vq,Feng:2010gw,Roszkowski:2017nbc,Schumann:2019eaa,
Lin:2019uvt,Arcadi:2017kky} have been considered as the most coveted 
candidate for dark matter particle with mass 
roughly between tens of GeV to a few TeV, and sufficiently large 
(of the order of electroweak strength) interaction with 
SM particles. WIMPs provide the observed relic 
abundance $(\Omega h^2 \sim 0.1198)$ via the well-known thermal 
freeze-out mechanism. Non observation of any experimental signal 
\cite{PandaX-II:2016vec, XENON:2017vdw, LUX:2016ggv,PICO:2019vsc,
AMS:2013fma,Buckley:2013bha,Gaskins:2016cha,Fermi-LAT:2016uux,MAGIC:2016xys,
Bringmann:2012ez,Cirelli:2015gux,Kahlhoefer:2017dnp,Boveia:2018yeb} 
of the DM lead to severe constraints on the WIMP paradigm.

To circumvent these constraints 
on the WIMP scenario, an alternative framework called {\it freeze-in} mechanism  has been
proposed. In this scenario the dark matter is a feebly interacting massive particle 
(FIMP) having highly suppressed interaction strength $\lesssim {\cal O}(10^{-12})$
with the SM sector. In the simplest scenario, it is assumed that the initial number
density of DM is either zero or negligibly small, and the observed relic abundance
is produced non-thermally either from annihilation or decay of SM 
particles in the early universe. The FIMP freezes in once the 
temperature drops below the dark matter mass and yields a fixed DM 
relic abundance that is observed at the present day 
\cite{Hall:2009bx,Konig:2016dzg,
Biswas:2016bfo,Biswas:2016iyh, Bernal:2017kxu,Borah:2018gjk}. 
FIMP having such a small coupling with the visible sector can 
trivially accommodate various null results of DM in different direct detection 
experiments such as Panda\cite{PandaX-II:2016vec}, XENON\cite{XENON:2017vdw},
LUX\cite{LUX:2016ggv}. However, FIMPs imprints can be traced via 
cosmological observations such as big bang 
nucleosynthesis (BBN), cosmic microwave background (CMB) or free
streaming length \cite{Heeck:2017xbu,Bae:2017dpt,Boulebnane:2017fxw,
Ballesteros:2020adh,DEramo:2020gpr,Decant:2021mhj,Li:2021okx, Ganguly:2022ujt}.

It would be very interesting to look for a minimal 
BSM paradigm where both the aforementioned sectors 
(non-zero neutrino mass and FIMP dark matter) are connected. 
In some particular scenario, such new interactions of neutrinos can 
have non-trivial implications in cosmological observations and
the precision era of cosmological measurements such as BBN or CMB 
provide us distinctive possibility for the indirect probe of those
hidden particles.

We know that one of the very important and precisely measured 
observable of the early Universe is the number of effective 
relativistic neutrino degrees 
of freedom or ${\rm N_{eff}}$ which can be changed in the presence of 
non-standard 
interactions of neutrinos. It is usually parameterised as, $
{\rm N_{eff}}\equiv {\left(\rho_{\rm rad}-\rho_{\gamma}\right)} /{\rho_{\nu_L}}$,
where $\rho_{\rm rad}$, $\rho_{\gamma}$ and $\rho_{\nu_{L}}$ are the total radiation energy density, energy density of photon and the energy density of single active neutrino species respectively. According to the recent data Planck 2018 
\cite{Planck:2018vyg}, at the time of CMB formation ${\rm N_{eff}^{CMB}=2.99^{+0.34}_{-0.33}}$ with 95$\%$ confidence level  
whereas the SM predicts it to be ${\rm N_{eff}^{SM}=3.045}$ \cite{Mangano:2005cc,Grohs:2015tfy,deSalas:2016ztq}.
The departure from 3, the number of neutrinos in the SM, is the consequence of 
various non-trivial effects like non-instantaneous neutrino decoupling, finite 
temperature QED corrections to the electromagnetic plasma and flavour 
oscillations of neutrinos. So, there are still some room to accommodate the 
contribution from the beyond SM physics. However, future generation 
CMB experiments like SPT-3G\cite{SPT-3G:2019sok}, 
CMB-IV\cite{CMB-S4:2016ple} are expected to attain a precision of $\rm{\Delta N_{eff}}\approx 0.06$ at 95\% confidence level. Thus any new contribution to the radiation energy density can be probed
very precisely which can constrain various BSM scenarios that 
produce light degrees of freedom and are in thermal equilibrium with 
the SM at early epoch of the evolution of our Universe.

On the other hand, different cosmological events such as decoupling of 
any relic species from the thermal bath or the non-thermal production of some species are sensitive to evolution history of the Universe. 
In the standard cosmological picture, it is assumed that after the end of the inflation, the energy density of the Universe is mostly 
radiation dominated. However, we only have precise information about the
thermal history of the Universe at the time of BBN ($T_{\rm BBN}\sim {\rm MeV}
$) and afterwards when the Universe was radiation dominated 
\cite{Kawasaki:2000en,Ichikawa:2005vw}. This allows us to consider 
the possibility that some non-standard species dominated significantly 
to the total energy budget of the Universe at early times $(T > T_{\rm BBN})$.
In that scenario, if the total energy density is dominated by some 
non-standard species, the Hubble parameter $(H)$ at any given 
temperature is always larger than the corresponding 
value of $H$ for 
the standard cosmology at the same temperature. Such a scenario with larger Hubble parameter is known as the fast expanding universe where at earlier time (higher temperature) the universe was matter dominated and eventually at some lower temperature  before BBN radiation density $(\rho_{\rm rad})$ takes over the energy  
density of the non-standard species ($\rho_{NS}$). One such possibility was discussed in \cite{DEramo:2017gpl} where $\rho_{NS}$ depends on the scale factor as $\sim a^{-(4+n)}$, where  $n>0$. Following the notation of \cite{DEramo:2017gpl}, this era of the
universe is identified by temperature $T_r$, where 
$\rho_{NS} (T_r) = \rho_{\rm rad} (T_r) $. Thus, the 
non-standard cosmological era correspond to the temperature 
regime where $ T > T_r$. In the limit, $n = 0$ corresponds to the standard radiation dominated cosmological picture. This may be realized by introducing a BSM scalar field $\Phi$ with equation of state(EOS) $p_{\Phi}=\omega \rho_{\Phi}$, where $p_{\Phi},\,\rho_{\Phi}$ denote the pressure and energy density of $\Phi$ respectively and $\omega\in{[-1,1]}$. The energy density  prior to BBN red-shifts as 
follows : 
\begin{eqnarray}
\rho_{\Phi}\propto a^{-(4+n)},
\label{eq:rphi}
\end{eqnarray}
where, $n= 3\omega-1 $. For $\omega>\frac{1}{3}$ this the energy density will fall faster than the radiation. This scenario have been studied in different context in the literature \cite{PhysRevD.55.1875,PhysRevD.28.1243, article,PhysRevD.37.3406,Copeland:1997et,Bernal:2018kcw}. We assume this $\Phi$ has negligible coupling with SM sector only so that the only effect it will have is in the expansion rate of the universe. As the new species red-shifts faster than the radiation. it's energy density will eventually become subdominant even without the presence of any decay. Several works reported the implications of  non-standard cosmological scenarios in context of WIMP relic density  calculation \cite{DEramo:2017gpl,Poulin:2019omz,Redmond:2017tja,
Hardy:2018bph,Barman:2021ifu,DEramo:2017ecx,Bernal:2018kcw}.
It is observed that if the thermal DM production occurs before 
$T_{\rm{BBN}}$ when the expansion rate of Universe was larger than radiation dominated(RD) 
universe, freeze out  happens at earlier time 
thus producing higher relic abundance 
\cite{Poulin:2019omz,Redmond:2017tja,Hardy:2018bph}. 
Thus one requires larger coupling of the DM with thermal bath particles 
to produce higher DM annihilation cross-section at late universe
so that it produces the relic density that matches with the observed one. 
Similar studies in the case of non-thermal production of DM have also been 
investigated \cite{DEramo:2017ecx,Co:2015pka,Bernal:2018kcw}. 
Various phenomenological implications of the 
non standard cosmology have been extensively studied by several groups
\cite{Berlin:2016vnh,Tenkanen:2016jic,Dror:2016rxc,Berlin:2016gtr}.

Motivated by this, we embark on a scenario where the SM particle 
content is augmented by an inert $SU(2)$ scalar doublet ($\eta$) and three SM 
gauge singlet fermions $(N_1,N_2,N_3)$ where all of them are odd under a 
Unbroken $\mathcal{Z}_2$ symmetry 
\cite{Ma:2006km,Borah:2018rca,Kitabayashi:2021hox,DeRomeri:2021yjo,
Escribano:2021ymx,Avila:2021mwg,Alvarez:2021otp,Escribano:2021wud,
Hundi:2022iva}. The striking feature of this model is the way it connects the 
origin of neutrino mass and DM. Neutrino mass can arise through  one loop 
radiative seesaw, whereas both $\eta^0$, the real component of the scalar
 doublet, and ${N}$ can be the DM candidate depending on their mass 
hierarchy. For scalar DM $(\eta^0)$, different 
studies\cite{Deshpande:1977rw,Barbieri:2006dq,LopezHonorez:2006gr,
Arhrib:2013ela} have shown that the correct relic density can be produced 
only in the high mass region ($M_{\eta^0}\gtrsim 525$ GeV). However, this 
scenario can change if we introduce another real gauge singlet scalar $S$, 
which is also odd under $\mathcal{Z}_2$ symmetry and mixes with $\eta^0$.
The immediate consequence of such non-trivial mixing between the singlet $(S)$
and $\eta^0$ is a newly formed scalar DM $(\eta_1)$ state as a linear 
superposition of $\eta^0$ and $S$ with suppressed gauge interaction compared
to the doublet scalar. As a result of this suppressed gauge coupling, 
the scalar DM can now have mass as low as $200$ GeV consistent with 
the observed relic density \cite{Beniwal:2020hjc,Farzan:2009ji, 
Ahriche:2016cio}.However, both in presence or absence of the singlet scalar, 
lightest of the singlet fermions $N_i$ can be a plausible thermal 
DM candidate due to their Yukawa interaction with new scalar doublet and 
the SM leptons. In all such cases, the model faces stringent constraints 
from different direct detection experiments. As discussed above, motivated by 
the null results of these experiments, here we study another version of 
scotogenic model where the DM is produced via non-thermal mechanism.  
In this analysis, instead of 
a scalar DM, the lightest singlet ${\mathcal Z}_2$ odd fermion $N_1$ plays 
the role of the DM, whereas the lightest neutral scalars $(\eta_1)$, 
which is the admixture of the real part of the doublet $(\eta^0)$ 
and the singlet $(S)$, is very long lived and decays 
to DM plus one neutrino at very late time (after neutrino decoupling).
If the decay happens at 
sufficiently low temperature, after the decoupling of active neutrinos 
from the thermal bath, it can significantly affect the total 
radiation energy density of the universe and contribute to the ${\rm N_{eff}}$.
While, calculating the amount of ${\rm \Delta{N_{eff}}}$, we realised that 
in standard cosmological scenario the remnant abundance of $\eta_1$ is not 
sufficient to produce detectable ${\rm \Delta{N_{eff}}}$ in the present 
experimental sensitivity. We show that the scenario can be significantly 
changed if our universe had gone through some non-standard expansion history.

This paper is organized as follows. The section \ref{sec:basic_set_up} 
is devoted for the brief discussion of the basic setup of the model and 
important interactions. The discussion on dark matter and 
${\rm \Delta N_{eff}}$ is presented in the section
section \ref{sec:neff}, while the section \ref{sec:res} contains our 
main numerical results. Finally, we conclude in section \ref{sec:concl}. 
\section{Basic set up}
\label{sec:basic_set_up}
The particle spectrum of this model contains a $SU(2)_{L}$ 
inert doublet scalar ($\eta$), a real singlet scalar ($S$) and three right-handed 
neutrinos ${N_i},(i=1,2,3)$ in addition to the SM particles. We impose an additional 
$\mathcal{Z}_2$ symmetry under which all the SM particles are even whereas the new fields are odd. In this prescription the lightest of these $\mathcal{Z}_2$ odd particles will be absolutely stable and be viable DM candidate. With these particles in 
hand, one can write down the following interaction Lagrangian:
\begin{eqnarray}
 \mathcal{L}_{\rm fermion}= y_{i \alpha } \overline{\ell^i_L} \tilde{\eta}
N_\alpha + \frac{1}{2}  M_{\alpha \beta } \overline{N_\alpha ^c} N_\beta  +h.c.,
 \label{eq:yuk}
\end{eqnarray}   
where $\ell^i_L$ is the SM left handed $SU(2)_{\rm L}$ lepton doublet,
$y_{i \alpha}$ is the lepton Yukawa coupling of flavours
$i = e, \mu,\tau $, $M_{\alpha \beta}$ is the symmetric Majorana mass matrix
and $\tilde{\eta}= i \sigma_2 \eta^*$.
The Yukawa interaction, in particular $y_{i1}$ term in equation 
\eqref{eq:yuk} plays the most important role in the dark matter 
phenomenology discussed later in this paper. The scalar potential of 
the model $V(\phi, \eta, S)$, followed by its minimization condition and 
relevant mass and coupling parameters are shown in Appendix \ref{apps}.

After the electroweak symmetry breaking, two neutral physical eigen states 
$\eta_1 $ and $\eta_2$ can be expressed as the linear combination of the 
weak basis $\eta^0$ and $S$ as:
\begin{eqnarray}
\eta_1 &=& \cos \theta\, \eta_0 - \sin \theta\, S\\
\eta_2 &=& \sin \theta\, \eta_0 + \cos \theta\, S
\label{diag}
\end{eqnarray}
where $\theta$ is the neutral CP-even scalar mixing angle. It is obvious
that for $\theta = 0$, $\eta_1 (\eta_2)$ doublet(singlet) dominated 
and vice-versa for $\theta = \pi/2 $.
The following parameters describe the scalar sector of this model 
(see Appendix \ref{apps} for details):
\begin{equation}
M_{\eta_1},\ M_{\eta_2},\ M_{A^0},\ M_{\eta^+},\ \lambda_{\eta},
\lambda_S,\ \lambda_{\phi S},\ \lambda_{\eta S},\ \lambda_3 ,\ 
\sin\theta 
\end{equation}
In addition to these we also have three right handed heavy neutrino 
masses, $M_{N_{1,2,3}}$. We use the following mass ordering and 
importance of this particular mass pattern in the context of our 
phenomenology will be discussed shortly. 
\begin{equation}
{M_{N_3},\, M_{N_2},M_{\eta_2} > M_{\eta_1}> M_{N_1}}.
\end{equation}
This mass pattern implies that the lightest $\mathcal {Z}_2$ 
odd fermion $N_1$ is a suitable candidate for the 
DM having the Yukawa coupling $y_{i1}$ that features
in the production of $N_1$ from the decay of $\eta_1$.
\begin{table}[h]
\begin{tabular}{ |c|c|c|c| }
 \hline
   & $M_{\eta^{\pm}}$ (GeV) & $M_{A^0}$ (GeV)  & $M_{\eta_2}$ (GeV) \\
   \hline
 BP-1 & 180 & 250 & 400 \\
  \hline
BP-2 & 150 & 200 & 400 \\
 \hline
 BP-3 & 180 & 220 & 500\\
 \hline
\end{tabular}
\caption{Values of heavy scalar masses for 3 benchmark points.}
\label{tab:bps}
\end{table}
To reveal the implications of heavier scalars in our analysis we 
consider three distinct values of $M_{\eta^\pm}, M_{A^0}$ 
and $M_{\eta_2}$ as represented by three benchmark points 
BP-1, BP-2 and BP-3 in Table \ref{tab:bps}. 
 
There is an interesting manifestation of relative mass-splittings among
$M_{\eta_1}$, $M_{\eta_2}, M_{A^0} $ on the dark
matter phenomenology as well as on constraining the model parameter space 
from the electroweak precision data. 
As far as heavier neutrino masses $M_{N_{2,3}} $ are concerned, 
we set them at ${\cal O}(1)$ TeV throughout this analysis 
so that neutrino masses can be generated radiatively 
in the right ball park with ${\cal O}(1)$ Yukawa couplings. 

Now, a short discussion on the production 
mechanism of aforesaid heavy particles 
and how they remain in
thermal equilibrium at early universe is called for. 
Both $\eta_1 $ and $\eta_2$ can be produced in thermal bath 
of the early universe through their interactions with 
the SM gauge bosons and Higgs boson.
Heavy neutrinos ${N_{i}}$ being gauge singlet can interact with 
thermal plasma only through $\eta$ and their presence in the 
thermal bath solely depend on the Yukawa interactions as shown 
in equation \eqref{eq:yuk}. The DM 
$N_1$ is produced non-thermally from the decay of both $\eta_1$ and 
$\eta_2$ and the respective decay rates depend on the Yukawa 
coupling $y_{i1}$ and the corresponding scalar mixing angles 
$\cos\theta$ and $\sin\theta$. For our choice of benchmark points 
(see Table \ref{tab:bps}) $\eta_2$ mostly decays to $W^\pm \eta^\mp$ or 
$Z A^0$ pairs through gauge interactions, leaving negligible contribution 
towards $N_1$ production via Yukawa coupling. Hence, for all practical 
purposes, $\Gamma (\eta_2 \to \nu + N_1) \simeq 0 $.
\begin{figure}[tbh!]
 \includegraphics[scale=0.4]{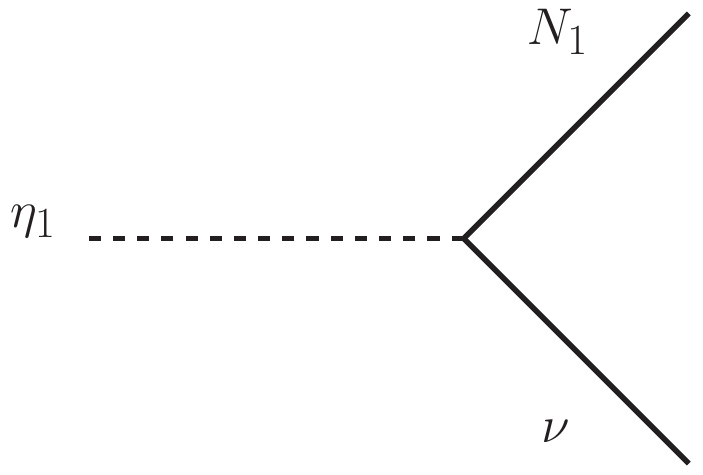} 
 \caption{Feynman diagram corresponding to the decay of $\eta_1$.}
\label{fig:decay}
\end{figure}
Thus for our choice of $\eta_1$ and $N_1$ masses, $N_1$ will be produced in 
association with active neutrino from the decay of $\eta_1$ 
(see Fig.\ref{fig:decay}) with $100\%$ 
branching ratio and the corresponding decay width for one neutrino generation 
can be written as 
\begin{eqnarray}
\Gamma_{\eta_1\rightarrow \overline{\nu}N_1}=\dfrac{y_{N_1}^2 \cos^2{\theta}\,M_{\eta_1} }{16\, \pi}  \left(1 - \dfrac{ M_{N_1}^2}{ M_{\eta_1}^2} \right)^2,
\label{eq:decay}
\end{eqnarray} 
where, we denote the Yukawa coupling $y_{i1}$ as $y_{N_1}$ and will use
this notation in rest of the paper. 
From the functional dependence of $\eta_1$ decay width 
(equation \eqref{eq:decay}) on $y_{N_1}$ 
it is clear that for the late lime production of $N_1$ from the
decay of $\eta_1$ via freeze-in mechanism, the Yukawa coupling $(y_{N_1})$
has to be extremely weak unlike for other two heavy neutrinos 
$(N_2~\&~N_3)$ that have $\mathcal{O}(1)$ Yukawa interactions with $\eta$.

In the presence of such large Yukawa couplings, both $N_2$ and $N_3$ will be produced in thermal equilibrium in the early epoch of the universe and will decay to other lighter $\mathcal{Z}_2$ odd particles ($\eta_1$, $\eta_2$, $\eta^\pm$, and $A^0$) after their decoupling from the thermal plasma. Hence, these decays would have no impact in the relic abundance of $N_1$.
So far, we were silent about the production of neutrino in association with 
$N_1$ from decay of long-lived $\eta_1$ and its impact on the dynamics 
of cosmology. The Yukawa coupling $(y_{N_1})$ is such that 
the decay $\eta_1 \to N_1 + \nu$ mostly happens after neutrinos 
decouple from the thermal bath and the decay must also be completed before 
CMB formation $(T\approx 1~{\rm eV})$ so that produced neutrinos in this 
mechanism have very intriguing implications in the observation of the CMB 
radiation.
To fulfill this condition of $\eta_1$ decay,
$y_{N_1}$ can not take any arbitrary value, rather 
it should be in the range  $(\sim 10^{-15} - 10^{-12})$ as considered 
in our analysis. As a result, this decay will inject entropy to the neutrino 
sector only and will increase the total radiation energy density of the 
universe at that epoch. However, any significant increment of total radiation 
energy density during CMB formation will directly impact 
${\rm \Delta N_{eff}}$, as discussed earlier and can be observed in 
different experiments. The same decay will also set the observed relic 
density of DM in today's universe. Hence, the DM mass will decide the amount of 
energy get transferred to the neutrino sector and can directly be 
related to $\rm{\Delta N_{eff}}$. The most important parameters of our 
discussion are $M_{\eta_1},\, \lambda_3,\ \lambda_{\phi S},\ \sin{\theta},
\ M_{N_1}\, \text{and}\, y_{N_1}$. Among these parameters,
${M_{\eta_1}}$,  $\lambda_3$ and $\lambda_{\phi S}$ decide the annihilation cross-sections of 
$\eta_1$, $\sin{\theta}$ decides mixing of the CP even real scalars 
$\eta^0$ and $S$. 
At the completion of $\eta_1$ decay, the final abundance of physical state
$\eta_1$ gets distributed into DM abundance and active neutrino 
energy density depending on $M_{N_1}$ and $y_{N_1}$, thus providing 
a connection between the DM mass (${ M_{N_1}}$) and ${\rm \Delta N_{eff}}$, 
and we will explore this in our current endeavour. As we prefer an 
enhanced co-moving number density of $\eta_1$ 
in the mass regime $M_{\eta_1}\gsim{ 60}$ GeV, various 
co-annihilation processes between dark sector particles must be suppressed
in order to avoid any additional enhancement of effective cross-section 
of $\eta_1$ as this would lead to lower abundance of comoving number 
density of $\eta_1$.
The lower yield of $\eta_1$ in turn produce lower neutrino number density and 
this may not be sufficient enough to induce observable effect on 
$\Delta N_{\rm eff}$. Hence, we suppress the above 
co-annihilation processes by increasing mass splittings between $M_{\eta_1}$ 
and other relevant heavy scalar particles of dark sector. This justifies our 
choice of associated heavy scalar masses for three benchmarks 
(BP-1, BP-2 \& BP-3) as shown in Table \ref{tab:bps}.

\section{ Freeze in DM and ${\rm \mathbf{\Delta N_{eff}}}$ at CMB}
\label{sec:neff}
Following our detailed discussions in previous sections, hereby we
address the issue of dark matter $(N_1)$
abundance created by the late time $(\tau_{\eta_1} > t_{\rm BBN})$
decay of $\eta_1 \to N_1 + \nu $. Since $N_1$ is the dark matter particle, it must satisfy the observed
relic abundance at present time and to estimate it one has to solve the following two coupled Boltzmann equations
that showcase the evolution of comoving number densities $Y_{\eta_1}$
and $Y_{N_1}$ corresponding to $\eta_1$ and $N_1$ respectively
with the temperature of the universe:
\begin{eqnarray}
\nonumber
\dfrac{dY_{\eta_1}}{dx} &=& -\dfrac{s}{H(x) x}\left(1-\frac{1}{3} \frac{d \ln g_{s}(x)}{d\ln x}\right) \left< \sigma v \right>_{\rm eff} (Y_{\eta_1}^2 -(Y_{\eta_1}^{eq})^2 )\\ &&-\dfrac{\left< \Gamma_{\eta_1\rightarrow N_1 \nu}  \right>}{H(x) x} \left(1-\frac{1}{3} \frac{d \ln g_{s}(x)}{d\ln x}\right)Y_{\eta_1},\label{eq:eta0}\\
\dfrac{dY_{\rm N_1}}{dx} &=& \dfrac{\left< \Gamma_{\eta_1\rightarrow N_1 \nu}  \right>}{H(x) x} \left(1-\frac{1}{3} \frac{d \ln g_{s}(x)}{d\ln x}\right)Y_{\eta_1}.
\label{eq:DM}
\end{eqnarray}
where $x=M_{sc}/T$ is a dimensionless variable with
$M_{sc}$ is some arbitrary mass scale which doesn't affect the
analysis and we consider it to be 100 GeV. Moreover, ${ Y_{\eta_1}^{eq}}$
is the equilibrium co-moving number density of $\eta_1$,
$g_s(x)$ and $H(x)$ represent the effective relativistic 
degrees of freedom related to the entropy density and the
expansion rate of the universe respectively. The thermal average
of effective annihilation cross-section of $\eta_1$ to the bath particles
is denoted by $\left< \sigma v\right>_{\rm{eff}}$. The entropy density
$s$ and $Y_i$'s are related as $Y_i=\frac{n_i}{s}$ where $n_i$'s are
the respective number densities. Finally $\left< \Gamma_{\eta_1\rightarrow N_1 \nu} \right>$ 
denote the thermal average of the decay width given
in equation \eqref{eq:decay}. While doing our numerical calculation, we take into
account all three active neutrinos in $\left < \Gamma_{\eta_1 \to N_1 \nu} \right > $. 

The evolution equation of the co-moving number density of $\eta_1$ is represented by 
equation (\ref{eq:eta0}). The first term on the right hand side
of this equation corresponds to the self annihilation of $\eta_1$ into the SM sector and 
vice versa, which keep $\eta_1$ in thermal equilibrium. However, 
in the presence of a tiny Yukawa coupling, $\eta_1$ slowly decays into 
$N_1 + \nu$, thus diluting its number density. This feature is reflected 
in the second term of equation (\ref{eq:eta0}).
Once the DM $(N_1)$ is produced in the above decay channel, its thermal 
evolution is governed by equation (\ref{eq:DM}). Note that in the absence of the Yukawa 
interaction, $\eta_1$ becomes stable and plays the role of the DM, having no effect on 
$\Delta N_{\rm eff}$, thus not considered in this analysis.  

We will now discuss the phenomenological consequence of late-time decay of $\eta_1$ into 
DM and neutrinos which is the motivation of this work. The sufficient production of active 
neutrinos after it decouples from the thermal bath ($T\lsim2$ MeV) can hugely affect 
the total radiation energy density of the universe at that time and will 
finally increase ${\rm \Delta N_{eff}}$. The effective number of relativistic neutrinos at the 
time of CMB can be written as: 
\begin{eqnarray}
N_{\rm eff}^{\rm CMB} = \frac{8}{7} \left(\frac{11}{4} \right)^{4/3} 
\frac{\rho_{\nu}}{\rho_{\gamma}}\Bigg|_{\rm T=T_{CMB}}.
\end{eqnarray}
where $\rm{\rho_\nu}$  and $\rm{\rho_\gamma}$ are the energy densities of neutrino and photon
 respectively. The production of $\nu$s from some external source will increase its energy 
density and we parameterize the deviation from the SM value in the following way: 
\begin{eqnarray}
\frac{N_{\rm eff}^{\prime}}{N_{\rm eff}^{\rm SM}} = \frac{\rho_{\nu}^\prime}{\rho_{\nu}^{\rm SM}}\Bigg|_{\rm T=T_{CMB}},
\end{eqnarray}
where $\rm{\rho_\nu^\prime}$ is the total energy density of neutrinos, i.e. the sum of the 
SM contribution (${\rho_\nu^{\rm SM}}$) and the non-thermal contribution 
$(\rho_\nu^{extra})$ coming from the decay of $\eta_1$. Hence,  
${\rm \Delta N_{eff}}$ can be expressed as follows,  
\begin{equation}
\Delta N_{\rm eff} = \left(\frac{\rho^\prime_\nu}{\rho^{\rm SM}_\nu} -1 
\right)\,N_{\rm eff}^{\rm SM} \Bigg|_{\rm T=T_{CMB}}
\label{eq:delN}
\end{equation}
To know the temperature evolution of the 
total neutrino energy density ${\rm \rho_{\nu}^\prime}$ after 
the decay of $\eta_1$, we need to solve the following Boltzmann equation:
\begin{eqnarray}
\frac{d \rho'_\nu}{d x} &=& - \dfrac{4\,\beta (T) \,\rho'_{\nu}}{x} + \frac{1}{x H (x)}\left< E \Gamma\right>_{\eta_1 \rightarrow N_{1}\nu} Y_{\eta_1} s,
\label{eq:rhonupr} 
\end{eqnarray}
where $\beta (T) $ shows the variation of $g_s(T)$ with T and is defined as:
\begin{equation}
\beta (T)= 1+ \frac{1}{3}\frac{T}{g_{s}(T)} \frac{d\,g_{s}(T)}{dT},
\end{equation}
and {$<E \Gamma>_{\eta_1 \rightarrow N_{1}\nu}$}, term associated with the thermal average 
of energy density transferred to neutrino sector 
from $\eta_1$ decay is defined as \cite{Biswas:2021kio}: 
\begin{eqnarray}
{\small \left< E \Gamma\right>_{\eta_1 \rightarrow N_{1}\nu}=  \frac{|\mathcal{M}|_{\eta_1 \rightarrow N_{1}\nu}^2}{32\pi} \frac{{\left(m_{\eta_1}^2-m_{N_1}^2\right)}}{m_{\eta_1}^2} \left(1-\frac{m_{N_1}^2}{m_{\eta_1}^2}\right)}.
\label{eq:rhonu}
\end{eqnarray}
The first term in the right hand side of equation \eqref{eq:rhonupr} shows the 
dilution of $\rho_{\nu}'$ due to expansion of the universe, whereas the 
second term shows the enhancement of $\rho_{\nu}'$ with $x$ 
after the decay of $\eta_1$.
The evolution of $\rm{\rho_{\nu}^{SM}}$ after neutrinos decouple from the 
thermal bath can be easily obtained by setting the term proportional 
to $Y_{\eta_1}$ of the of equation \eqref{eq:rhonupr} to be zero 
$(\rho^\prime_\nu = \rho^{\rm SM}_{\nu})$ which dictates the dilution of 
energy density due to expansion only.

From  equation \eqref{eq:rhonupr} it is well understood that the total 
energy density of neutrinos $(\rho_{\nu^\prime} = 
{\rho_\nu^{\rm SM}} + \rho_\nu^{extra})$ is decided by the co-moving 
number density ($Y_{\eta_1}$) of $\eta_1$ after it freezes out.  
The freeze-out abundance of $\eta_1$ depends on its interaction with the 
bath particle and the expansion rate of the universe. Freeze-out of $\eta_1$ 
occurs at the temperature 
where the expansion rate, $H$ is greater than the interaction 
rate($\left<\sigma v \right>$).
$Y_{\eta_1}$ decreases if the freeze-out happens 
at late time equivalently at lower temperature. 
Thus making the Hubble as important quantity 
that determines the abundance $Y_{\eta_1}$.

In the standard cosmology where it is assumed that, 
universe at the time of DM freeze out is radiation dominated and 
the corresponding Hubble parameter $(H)$ is defined as:
\begin{equation}
H(T)=\sqrt{\dfrac{8 \pi \, G\, \rho_{\rm rad}(T)}{3}},
\label{eq:hubble}
\end{equation}
where G is the gravitational constant and $\rho_{\rm{rad}}(T)$ is the 
radiation energy density which scales as $\sim T^4$. 
 It turns out that 
for the range of $M_{\eta_1}$ considered in our analysis the standard radiation
dominated universe gives $\Delta N_{\rm eff}$ far below the current 
experimental sensitivity. This is due to the sizable 
interactions of $\eta_1$ with the SM bath, in other words large 
$\left<\sigma v \right>_{\rm eff} $, that keeps $\eta_1$ in 
thermal equilibrium for sufficiently longer duration. Such a large annihilation cross-section
of $\eta_1$ naturally produces low freeze-out abundance $Y_{\eta_1}$
as $Y_{\rm F.O} \propto 1/ \left<\sigma v\right>_{\rm eff}$. 
Hence, the number density of neutrinos produced from such a 
low yield $\eta_1$ is not sufficiently large enough to make any 
significant changes in $\Delta N_{\rm eff}$ that can be measured with 
current experimental precision. Interestingly, the situation changes 
drastically if we consider some non-standard species $\Phi$ 
that dominate the total energy budget of 
universe at early epoch where the universe goes through faster 
expansion at the time of $\eta_1$ freeze-out. Here one assumes 
that in the pre-BBN era, the energy density of the universe 
receives contributions from both the radiation as well as 
a new species $\Phi$. The energy 
density of $\Phi$ scales as $\sim a^{-(4+n)}$ for $n>0$ and can be 
rewritten in the following form :
\begin{eqnarray}
\rho_{\Phi}(T) = \rho_{\Phi}(T_r) \left[ \frac{g_{s}(T)}{g_{s}(T_r)}
\right]^{(4+n)/3} \left (\frac{T}{T_r} \right )^{(4+n)},
\end{eqnarray}
where, $g_{s}$ is the effective degrees of freedom contributing to the 
entropy density. We consider $T_r$ as the temperature where 
$\rho_{\Phi}  = \rho_{\rm rad}$. Thus, using the entropy 
conservation law, one can express the total energy density 
$(\rho (T) = \rho_{\Phi}(T) + \rho_{\rm rad}(T)) $ at a given 
temperature $T$ in the following form \cite{DEramo:2017gpl}:
\begin{equation}
\rho(T)= \rho_{\rm {rad}}(T)\left[1+ \dfrac{g_{\rho}(T_{\rm{r}})}
{g_{\rho}(T)} \left(\dfrac{g_{s}(T)}{g_{s}(T_{\rm{r}})}\right)
^{\frac{4+n}{3}} \left(\frac{T}{T_{\rm{r}}} \right)^n \right],
\label{eq:fast}
\end{equation}
where, $g_{\rho}$ is the effective relativistic degrees of freedom. 
For $T>T_{\rm r}$ the energy budget of the universe is dominated by  $\Phi $. The Hubble parameter $H$ in 
equation \eqref{eq:hubble} is now determined 
by the total energy density $\rho(T)$ as shown in 
equation (\ref{eq:fast}) instead of only $\rho_{\rm rad}(T)$. 
Note that $n=0$ limit recaptures the standard cosmological picture.
Hence, $T_r $ and $n$ are the two 
important parameters that decide the expansion 
rate $H$. For $n> 0$ the expansion rate of the universe at a 
given temperature $T$ is always larger than the 
corresponding value in the standard radiation dominated $(n=0)$
scenario. As a result of this fast expansion, the 
condition $\left< \sigma v\right>_{\rm{eff}} 
< H(x) $ is achieved earlier and $\eta_1$ freezes out 
at temperature $T$ higher than than the corresponding temperature in
standard cosmological picture. With such an earlier freeze-out 
the abundance $Y_{\eta_1}$ is large enough to significantly 
increase the amount of $\Delta_{N_{\rm eff}}$ in our proposed 
model. However, one should be careful from potential impact
of above phenomena on the successful predictions of light element 
abundance by the BBN. If $T_r$ is close to BBN temperature 
$T_{\rm BBN}\sim 1~{\rm MeV}$, the universe starts to expand faster than 
the radiation dominated picture around $t_{\rm BBN}$ and this may 
modify the theoretical prediction for BBN abundances. To avoid this
$T_{\rm r}$ must satisfy the following condition \cite{DEramo:2017gpl}:
\begin{eqnarray}
T_{\rm r}\gtrsim (15.4)^{\frac{1}{n}}~{\rm MeV}. 
\label{eq:tbbn}
\end{eqnarray}

Upto this point,  we were silent about the nature of $\Phi$ or the essential potential which can give rise to expansion faster than usual RD universe and treated $n$ and $T_r$ as free parameters. Followed by the discussion in the introduction we assume $\Phi$ to be a scalar field which is minimally coupled to gravity with a
positive self-interacting scalar potential($V(\Phi)$).
The EOS parmeter ($\omega$) lies in the range $\omega\in[-1,1]$
i.e. $n\in [-4,2]$ depending on whether the potential energy $V(\Phi)$  or kinetic energy(KE) term  dominates \cite{PhysRevD.58.023503,PhysRevD.28.1243,PhysRevLett.79.4740}.
The former situation is realized if $\Phi$ is oscillating about the minimum of a positive potential \cite{PhysRevD.28.1243} 
and have been studied in different contexts \cite{Kallosh:2013hoa,Kallosh:2013yoa,STAROBINSKY198099,KOFMAN1985361}.
On the other hand, the scenario  where, the universe's energy density is dominated by the KE of the scalar field, gives rise to the later one($n=2$), which is often known as kination \cite{PhysRevD.55.1875,article,PhysRevD.37.3406,Copeland:1997et}. Such theories with $n=2$ are relaizations of quintessence fluids motivated to explain accelarated expansion of the universe 
     \cite{PhysRevLett.80.1582,Wetterich:1994bg,Sahni:1999gb}. However in this work, we are interested to study how fast expanding universe enhances the abundance $Y_{\phi}$ as well as  $\rm{\Delta N_{eff}}$ for any $n>0$.

In the next section, we will discuss comprehensive numerical analysis of 
relic density calculation along with $\rm{\Delta N_{eff}}$ and its 
phenomenological implications. While doing our numerical analysis 
we will vary parameters $T_r$ and $n$ such that they satisfy 
equation (\ref{eq:tbbn}).

\section{Numerical results}\label{sec:res}
In the previous section we have argued that if some non-standard matter 
field $\Phi$ dominates the total energy budget of the universe at early epoch,
the universe goes through a faster expansion and $\eta_1$ freezes 
out early with sufficiently large relic abundance.  From $\eta_1$
the dark matter $(N_1)$ and neutrinos are produced via freeze-in 
mechanism. Thus the produced number density of neutrinos are sufficiently
large enough to make a substantial new contribution to $\Delta N_{\rm eff}$ 
which can be verified in the current experiment. 
In this section we will scan our model parameter space to 
quantify this modified $\Delta N_{\rm eff}$. We will also 
show that for a given set of model parameters, $\Delta N_{\rm eff}$ is 
highly correlated with the dark matter mass $M_{N_1}$. Any direct experimental
verification of FIMP like dark matter is extremely challenging due its 
tiny coupling with SM sector. However, in this scenario, we find that 
the observed value of 
$\Delta N_{\rm eff}$ is strongly dependent on $M_{N_1}$ and 
one can utilize this observable as an experimental probe 
for FIMP like dark matter. The presence of additional scalars ($SU(2)$ doublet and a
singlet) and three generations of heavy neutrinos can have important
implications on various existing experimental data. Hence, to have
a phenomenologically consistent model, it is necessary to
carefully scrutinize aforementioned BSM scenario in the light of
those experimental data. In addition to these, mathematical consistency
of the scenario also demands that various model parameters must satisfy
certain theoretical conditions. However, for the brevity of the analysis we will
not discuss these here, nevertheless, further details can be seen in \cite{Beniwal:2020hjc} 
\begin{figure}[htb!]
\centering
\subfigure[\label{a}]{\includegraphics[scale=0.4]{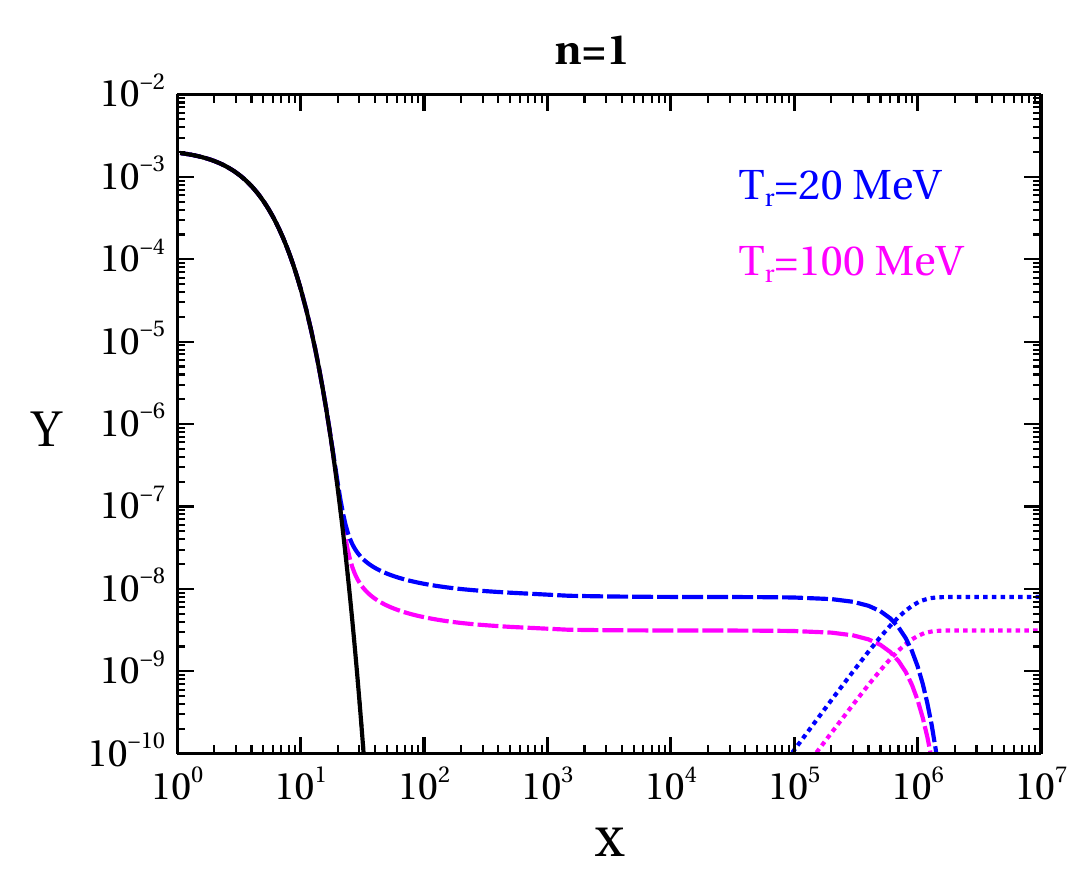} }
\subfigure[\label{b}]{\includegraphics[scale=0.4]{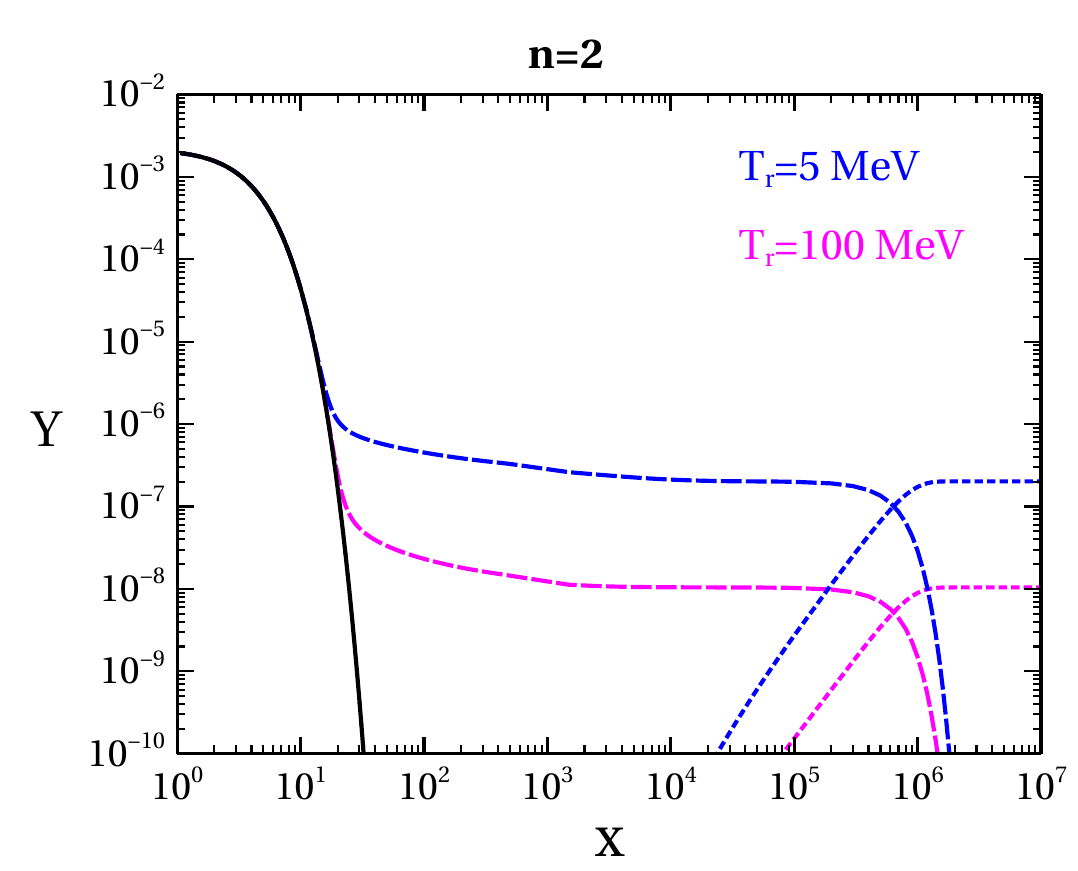}}
 \caption{Evolution of co-moving number densities of $\eta_1$ and $N_1$ as a 
function of dimensionless variable $x$ for $\lambda_{3}=10^{-3}$, 
$\lambda_{\phi S}=10^{-3}$, ${M_{N_1}}=10$ MeV, $M_{\eta_1}=65$ GeV, 
$y_{N_1}=10^{-12}$, $\sin{\theta}=0.9$. Solid, dashed and dotted lines
correspond to $Y_{\eta_1}^{eq}$, $Y_{\eta_1}$ and $Y_{N_1}$ respectively. 
Co-moving number densities are shown for different values of $T_r$ in 
panel $(a)$ for $n=1$, and in panel $(b)$ for $n=2$.}
\label{fig:DM}
\end{figure}
We will first discuss the full numerical solutions to the 
Boltzmann equations corresponding to $Y_{\eta_1}$ (Eq.~\ref{eq:eta0})
and $Y_{N_1}$ (Eq.~\ref{eq:DM}) respectively. For this analysis, we first 
implement the interactions, mass and mixings of the model in 
\texttt{FeynRules}~\cite{Alloul:2013bka}, that generate required 
\texttt{CALCHEP}\cite{Belyaev:2012qa} model files for \texttt{micrOMEGAs}
\cite{Belanger:2014vza} to calculate thermally averaged cross-section 
$<\sigma v>_{\rm eff}$.
To showcase the behaviour of $Y_{\eta_1}$ and $Y_{N_1}$ with temperature
$T$ we consider $M_{\eta_1} = 65 $ GeV\footnote{For this value of $M_{\eta_1}$, 
$h \to \eta_1 \eta_1$ is kinematically forbidden and hence no constraints from
${\rm BR}_h \to {\rm inv} $.}, $M_{N_1} = 10 $ MeV, 
$\lambda_3 = 10^{-3}, \lambda_{\phi S} = 10^{-3}, \sin\theta =0.9 $ and
the Yukawa coupling $y_{N_1} = 10^{-12}$. It is worth pointing out that those
parameters are consistent with all theoretical and experimental constraints
discussed earlier. We have two additional parameters $T_r$ and $n$ that fix 
the cosmological framework of our present scenario. 
The co-moving number densities $Y_{\eta_1}$ and $Y_{N_1}$ 
for $\eta_1$ and $N_1$ are plotted as a function of $x$  
for $n=1$ (Fig.\ref{fig:DM}$(a)$) and $n=2$ (Fig.\ref{fig:DM}$(b)$) 
respectively. 

In Fig.\ref{fig:DM}$(a)$ blue and magenta lines correspond to 
$T_r$ = 20~MeV and 100~MeV, while in Fig.\ref{fig:DM}$(b)$ the corresponding
two colored lines represent $T_r = $ 5~MeV and 100~MeV respectively. 
From equation \eqref{eq:hubble} and equation \eqref{eq:fast} 
one can find that in the presence of an extra contribution to the
energy density of the universe, the Hubble parameter $H$ goes like 
$\sim T^2 \left(T/T_r\right)^{n/2},~~(T >> T_r)$ and this explains 
why the expansion rate of the universe increases with the increase 
in $n$(for $n>0$) which ultimately leads to an earlier freeze-out 
of $\eta_1$ with higher abundance. 
This feature of fast expanding universe is nicely evinced 
in both Fig.\ref{fig:DM}$(a)$ and Fig.\ref{fig:DM} $(b)$, where, 
co-moving number densities for $\eta_1$ as well as $N_1$ are higher 
for $n=2 $ compared to $n=1$ for a fixed $T_r=100$ MeV(magenta lines). 
It is clearly evident from Fig.\ref{fig:DM} that ${\eta_1}$ 
decouples from the thermal bath first and then decays 
into $N_1 + \nu$ at some later time which basically
increases the dark matter co-moving number density $Y_{N_1}$. 
Besides, one can also notice from the Fig.\ref{fig:DM} that, 
for a fixed $n$, $Y_{\eta_1}$ decreases with the increase in 
$T_{r}$ and this can be once again traced back to the parametric 
dependence of the Hubble $H$ on $T_{r}$ and $n$ as mentioned earlier.
In summary, both $Y_{\eta_1}$ and $Y_{N_1}$ increase with increase in 
$n$ for a given $T_r$, while they decrease with increase in $T_r$ for 
any value of $n \geq 1 $. Interestingly both these observations can be 
interpreted in terms of the modified Hubble parameter $H(T)$ 
in the fast expanding universe.

\begin{figure}[tbh!]
 \includegraphics[scale=0.38]{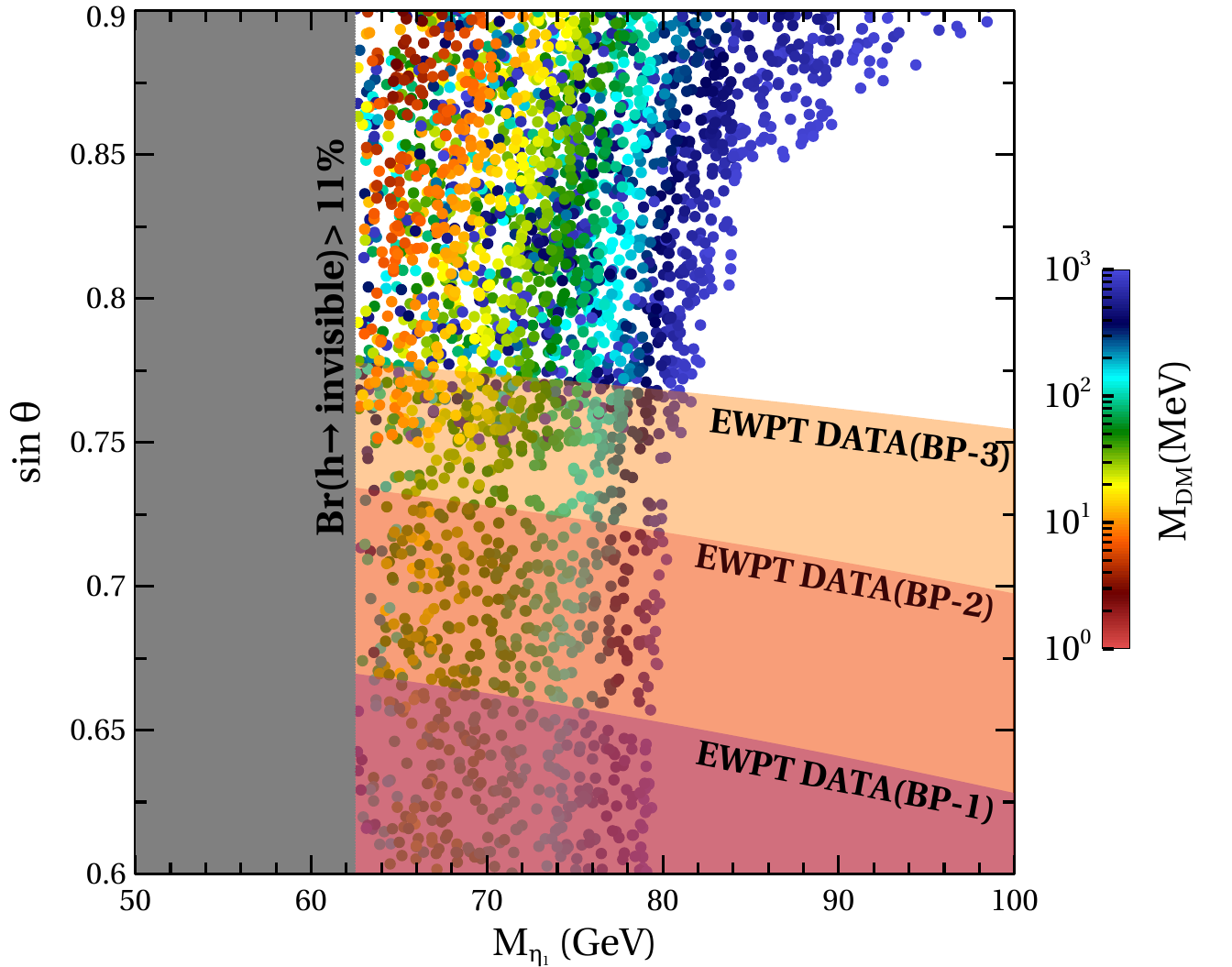} 
 \caption{The allowed parameter space from the electroweak precision test (EWPT) data 
and Higgs' invisible decay width constraints in ${\rm \sin{\theta}-M_{\eta_1}}$ plane for  $10^{-6}\leq\lambda_{3}\leq10^{-2}$, $10^{-6}\leq\lambda_{\phi S}\leq10^{-2}$,$y_{N_{1}}=10^{-12}$ $n=2$, $T_r=5$ MeV. Here the color bar shows the variation of DM mass.}
\label{fig:DM:scan}
\end{figure}

After illustrating the importance of the fast expanding universe in the 
calculation of relic abundances for both $\eta_1$ and $N_1$ we now scan the
independent parameters of the model in the following range:  
\begin{eqnarray}
&& \lambda_3 \in [10^{-6} : 10^{-2}]\,, ~\lambda_{\phi S} \in [10^{-6} : 10^{-2}] \\
&& M_{\eta_1} \in [50~{\rm GeV} : 100~{\rm GeV}]\,, M_{N_1} \in [1~{\rm MeV} : 1000~{\rm MeV}]\\
&& \sin \theta \in [0.0 : 0.9] 
\end{eqnarray}
The purpose of this scan is to find a region in the multi-dimensional 
model parameter space that is allowed by both theortical and experimental 
constraints as well as satisfy the correct relic density  
\cite{Beniwal:2020hjc,Kannike:2012pe,PhysRevD.46.381,Haber:2010bw,ATLAS:2020kdi}. 
We will then use those allowed parameters to calculate the value 
of $\Delta N_{\rm eff}$ in our model that can be substantiated 
in the upcoming experiments. Throughout this analysis we fix the mass of 
the SM-like Higgs to 125 GeV. One can notice that we 
vary $\sin\theta $ in the above range $(0.0 - 0.9)$
so that we can capture the effect of both 
$SU(2)$ doublet $\eta^0$ and singlet $S$ scalars in the 
freeze-in production of the DM in $\eta_1 $ decay, where 
the heavy scalar $\eta_1$ becomes doublet (singlet) 
dominated for $\sin\theta = 0 (0.9)$. For this analysis, 
we fix $n=2$, $T_r = 5$ MeV and
the Yukawa coupling, $y_{N_1} = 10^{-12}$. Varying the Yukawa coupling in the range as mentioned  in section \ref{sec:basic_set_up} will have no impact on relic density because  $$\Omega_{\rm DM}h^2= 2.755\times 10^8 \, Y_{\rm DM}\, \frac{M_{\rm DM}}{\rm GeV}, $$ and $Y_{\rm DM}$ will always be same as $Y_{\eta_1}(x_{f.o.})$. At the end the parameter 
scan result in ${M_{\eta_1}}-\sin{\theta}$ plane 
is visible in Fig.\ref{fig:DM:scan}, where DM mass $M_{N_1}$ is 
represented by the color bar. As one can observe that a significant
region of the parameter space in Fig.\ref{fig:DM:scan} has been excluded by various 
theoretical and experimental constraints. The ${\rm Br}(h \to~{\rm inv})< 11\% $ excludes
region with $M_{\eta_1} < M_h/2 $ and this is the grey coloured verical patch marked as 
${\rm Br}(h \to {\rm inv}) > 11\%$ \cite{ATLAS:2020kdi}. The electroweak precision data (EWPD) through 
$S,T,U$ parameters serve another
crucial limit on the parameter space of this model. It is well known that larger the mass 
splitting between the components of $SU(2)$ doublet field, stronger is the EWPD constraints \cite{Beniwal:2020hjc}. 
Indeed, this is happening in the case of our three benchmark points shown in 
Table \ref{tab:bps}. The EWPD data excludes three diagonal bands 
corresponding to BP-1, BP-2 and BP-3 respectively in Fig.\ref{fig:DM:scan}. 
From this figure, it is clear that the BP-3 which has the 
largest mass gap between $M_{\eta_2}$ and $M_{\eta^\pm}$ attracts the 
strongest EWPD limit and it excludes 
$\sin\theta \lsim 0.6-0.775$ for $M_{\eta_1} \approx 62.5 $ GeV and 
$\sin\theta \lsim 0.6-0.75$ for $M_{\eta_1} =100 $ GeV. 
Thus, the overall allowed region is located at the upper quadrilateral 
part of the parameter space, with $\sin\theta \sim 0.775 - 0.9 $ and 
$M_{\eta_1} \sim 62.5~{\rm GeV} - 100~{\rm GeV}$. Another important 
outcome of our analysis is that for a fixed $\sin\theta $ any 
increase in $M_{N_1}$, also call for an increase in $M_{\eta_1}$ to 
satisfy the correct density and Fig.\ref{fig:DM:scan} perfectly corroborate 
our claim. However, one has to look for any physical processes that may 
imperil the effect of long-lived scalar on cosmic microwave background (CMB)
in singlet-doublet scotogenic model. In this scenario, the presence of other heavy scalars may lead to DM co-annihilation processes which may boost the $\left< \sigma v \right>_{\rm eff}$ and such enhanced $\left<\sigma v\right>_{\rm eff}$ ultimately suppress
$\eta_1$ abundance. The immediate consequence of this low yield $\eta_1$ 
is the tiny production of additional neutrino density $\rho^\prime_\nu$
that may not lead to any significant shift in $\Delta N_{\rm eff}$, thus 
spoiling the intention of this analysis. The most natural way to circumvent this situation, is to take other particles of the model very heavy compared to $M_{\eta_1}$ 
(large mass splittings) so that one can easily ignore the co-annihilation 
of $\eta_1$ with those heavier particles. To facilitate this in our analysis 
we choose three representative benchmark points (BP-1, BP-2 \& BP-3) as 
shown in Table~\ref{tab:bps}. 

\begin{figure}[h]
\centering
\subfigure[\label{2a}]{\includegraphics[scale=0.53]{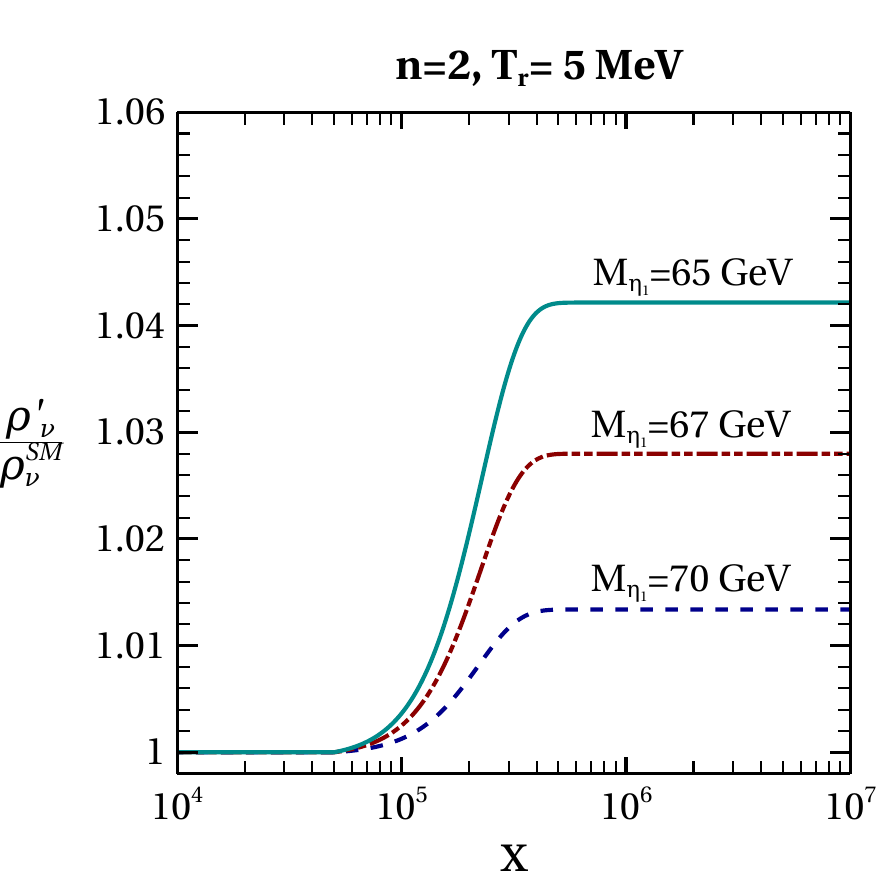}}
\subfigure[\label{2b}] {\includegraphics[scale=0.53]{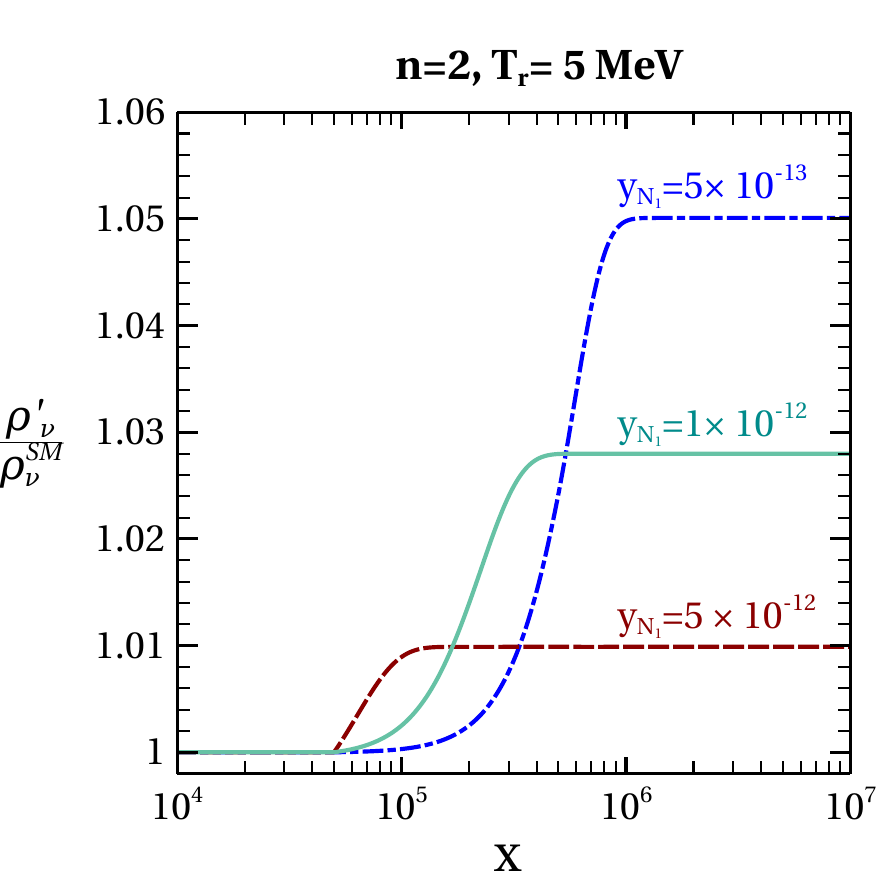}}
 \caption{Evolution of the ratio $\rm{\rho_{\nu}^\prime / \rho_{\nu}^{SM}}$ 
as a function of $x$ for $\lambda_{3}=10^{-3}$, 
$\lambda_{\phi S}=10^{-3}$, ${M_{N_1}}=10$ MeV, 
$\sin{\theta}=0.9$, $n=2$, $T_r=5$ MeV. We show our findings
for three different values of $\rm{M_{\eta_1}}$ for a fixed 
$y_{N_1}=10^{-12}$ in Fig.~\ref{2a} and for three different 
values of $y_{N_1}$ for constant $\rm{M_{\eta_1}}=67$ GeV in Fig.~\ref{2b}.}
\label{fig:Ratio}
\end{figure}
After having a suitable model parameter space consistent with
various constraints, we are now in a position to 
kickoff the numerical estimation of the ratio ${\rho_{\nu}'}/{\rho_{\nu}^{SM}}$.
For this we need to solve
equation \eqref{eq:eta0} and \eqref{eq:rhonupr} to 
evaluate the evolution of ${\rho_{\nu}'}/{\rho_{\nu}^{SM}}$ as
a function of the dimensionless variable $x$.
In Fig.\ref{2a} we display the dependence of ${\rho_{\nu}'}/{\rho_{\nu}^{SM}}$ as a function 
of $x$ for three different values of ${M_{\eta_1}}$ (65 GeV (solid cyan line), 
67 GeV (brown dot dashed line), 70 GeV (blue dashed line) 
where the other parameters are kept fixed ($\lambda_{3}=10^{-3}$, 
$\lambda_{\phi S}=10^{-3}$, ${M_{N_1}}=10$ MeV, $\sin{\theta}=0.9$, 
$n=2$, $T_r=5$ MeV) for a constant Yukawa coupling $y_{N_1}=10^{-12}$. 
From this figure (Fig.\ref{2a}) one can clearly discern that at very high 
temperature $T$, $\eta_1$ stays in thermal equilibrium for 
$M_{\eta_1} = 65 $ GeV and with no additional 
contribution to $\rho^\prime_\nu$ from $\eta_1$, thus 
the ratio ${\rho_{\nu}'}/{\rho_{\nu}^{SM}}$ is almost unity. However, as the 
universe starts cooling, the decay $\eta_1 \to N_1 + \nu$ 
also proceeds through tiny Yukawa coupling $y_{N_1}$ 
and it leads to new contributions to the neutrino energy
density. This additional contribution makes the ratio ${\rho_{\nu}'}/{\rho_{\nu}^{SM}}$ 
greater than one. The decay of $\eta_1$ continues
until its number density is completely converted into $N_1$ and $\nu$ number 
densities. At that point, no further neutrinos are generated and the 
${\rho_{\nu}'}/{\rho_{\nu}^{SM}}$ saturates. This feature is repeated for other two values 
of $M_{\eta_1}= 67$ GeV and $70$ GeV in Fig. \ref{2a} with a marked 
difference. For higher value of $M_{\eta_1}$ the thermally averaged 
cross-section $(\left<\sigma v\right>)_{\rm eff}$ of $\eta_1$ with SM particles 
increases due to larger phase space availability. This elevated cross-section
leads to the late time freeze out of $\eta_1$ with smaller 
abundance. Subsequently the late time decay of $\eta_1$ produces neutrinos
with suppressed energy density \cite{Beniwal:2020hjc}. 
From this analysis we conclude that for a fixed value of $x$, 
the ratio ${\rho_{\nu}'}/{\rho_{\nu}^{SM}}$ is the largest (smallest) 
for $M_{\eta_1} = 65 (70) $ GeV respectively. 

The Yukawa coupling $y_{N_1}$ also controls the temperature $T$ variation of 
the ratio ${\rho_{\nu}'}/{\rho_{\nu}^{SM}}$ for a fixed $\eta_1$ mass.
In Fig.\ref{2b}, we show such variation of ${\rho_{\nu}'}/{\rho_{\nu}^{SM}}$ with $x$
assuming three different values of $y_{N_1}$ 
($5 \times 10^{-12}$ (brown dashed line),
$1 \times 10^{-12}$(solid cyan line ), $5 \times 10^{-13}$
(blue dot dashed line) for $M_{\eta_1} = 67 $ GeV, while other parameters 
are same as in Fig.\ref{2a}. For a given mass of $\eta_1$ and $N_1$ and keeping
other model parameters fixed, the decay of $\eta_1 \to N_1 + \nu$ is 
completely determined by the Yukawa coupling $y_{N_1}$. Besides, 
larger the coupling $y_{N_1}$, faster is the decay rate of $\eta_1$ 
and vice-versa. For a given $y_{N_1}$, the ratio keeps on increasing 
with the expansion of the universe and ultimately 
saturates when the decay of $\eta_1$ is completed. For any 
higher $y_{N_1}$, the aforementioned decay of $\eta_1$ gets 
completed even at earlier time 
(equivalently at higher temperature $T$) and the energy injection to 
neutrino sector also completes at earlier epoch of the universe. 
This phenomena is distinctly noticeable 
from Fig.\ref{2b} that the ratio saturates earlier with increase 
in $y_{N_1}$. As the higher value of Yukawa coupling forces the $\eta_1$
decay to be completed at earlier time, the new contribution to neutrino energy 
density $\rm{\rho_{\nu}^\prime}$ from this decay gets diluted more due 
to the expansion of the universe. However with lower Yukawa coupling 
$y_{N_1} = 5\times 10^{-13}$, the decay starts later and the 
corresponding energy injection to neutrino sector also gets completed 
at later time (lower temperature) hence gets less diluted due to the 
expansion of the universe. Consequently,
one gets a higher value of ${\rho_{\nu}'}/{\rho_{\nu}^{SM}}$ at lower temperature 
($x\approx 10^7$) as depicted in Fig. \ref{2b}. 
\begin{figure}[tbh!]
 \includegraphics[scale=0.4]{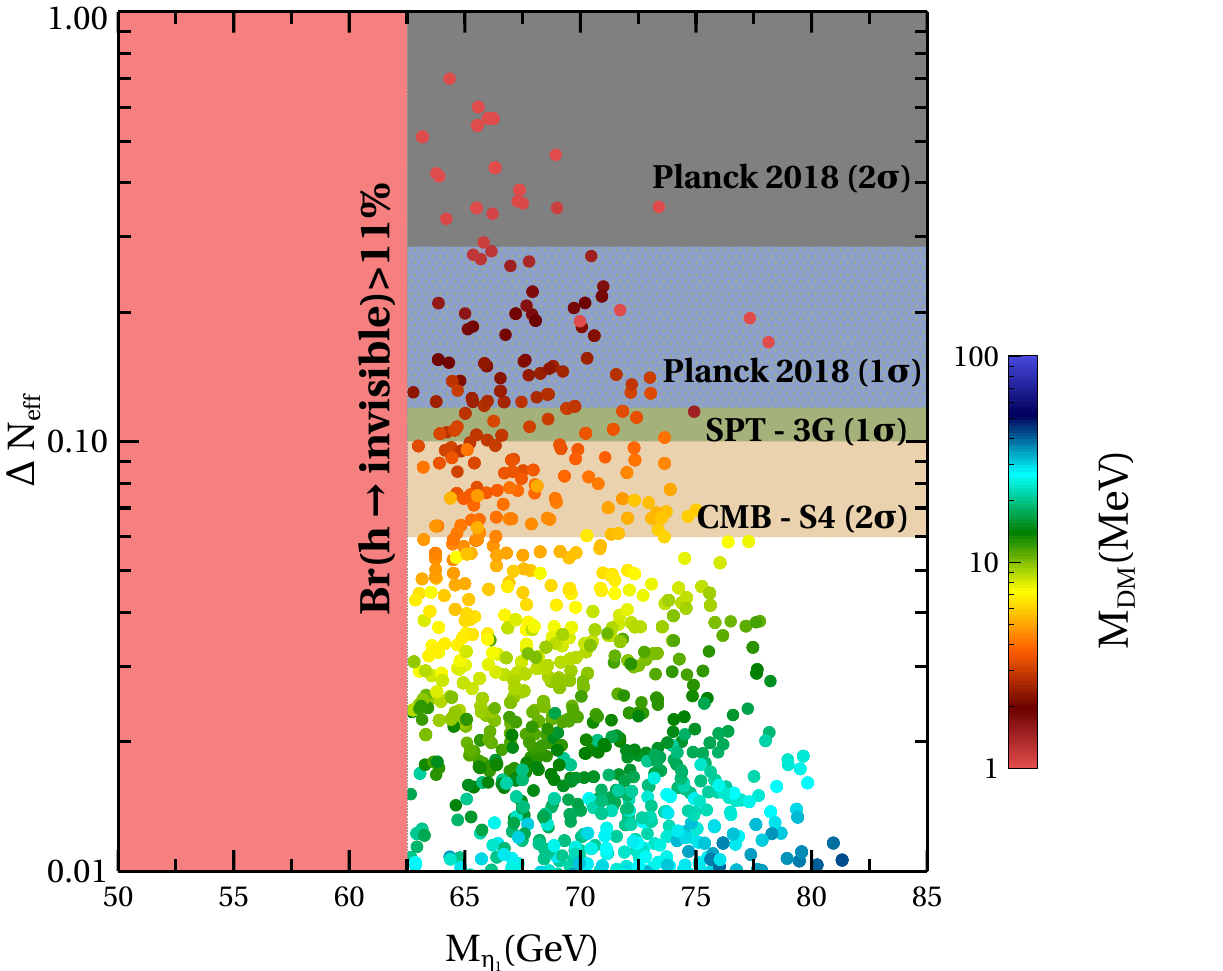}
 \caption{Variation of ${\rm \Delta N_{eff}}$ with $M_{\eta_1}$ for ${M_{A^0}}=250$ GeV, ${M_{\eta^+}}=180$ GeV, ${M_{\eta_2}}=400$ GeV, $10^{-6}\leq\lambda_{3}\leq10^{-2}$, $10^{-6}\leq\lambda_{\phi S}\leq10^{-2}$,$y_{N_1}=10^{-12}$ and  $n=2$, $T_r=5$ MeV. The color bar in the plot represents the DM mass. }
\label{fig:scan:2}
\end{figure}

After highlighting the variation of ${\rho_{\nu}'}/{\rho_{\nu}^{SM}}$ with various model parameters, we
now use ${\rho_{\nu}'}/{\rho_{\nu}^{SM}}$ to calculate the $\rm{\Delta N_{eff}}$ by scanning over 
those model parameters and the corresponding upshot is displayed 
in $\Delta N_{\rm eff}$ vs $M_{\eta_1}$ plane in Fig.\ref{fig:scan:2}. 
The color bar represents the value of $\rm{M_{N_1}}$ in the range 
$1~{\rm MeV} - 100~{\rm MeV}$.  
Here we consider BP-1 as the maximum allowed region from EWPD 
constraints correspond to this particular benchmark point.
Similarly other parameter
space points satisfy the correct relic density as well as are 
consistent with other theoretical and experimental
constraints. 
In the calculation of $\Delta N_{\rm eff}$, we vary the DM mass $M_{N_1}$ 
from $1$~MeV to $100$~MeV and it is conspicuous from Fig.\ref{fig:scan:2}, 
that with increase in $M_{N_1}$, the numerical value of 
$\rm{\Delta N_{eff}}$ decreases. This is understandable as 
a higher value of $M_{N_1}$ requires lower abundance $Y_{N_1}$ so that  
the correct relic density($\Omega h^2\sim Y_{N_1}M_{N_1}$) is achieved.
In our earlier discussions, we have shown that the evolution of $N_1$ abundance 
is governed by $Y_{\eta_1}$(Fig. \ref{a}), thus, lower $Y_{N_1}$ also 
corresponds to smaller $\eta_1$ abundance $Y_{\eta_1}$. 
With this lower abundance of $\eta_1$ lesser energy gets transferred to neutrino sector, leading to smaller value of $\rm{\Delta N_{eff}}$. For this scan we consider fixed $y_{N_1}=10^{-12}$. We may decrease $y_{N_1}$ that would lead an increase in $\rm{\Delta N_{eff}}$ as understood from Fig.\ref{2b}, and will make the scenario more viable to the observations. However the DM phenomenology will remain same. We also show the 
exclusion limit of $\rm{\Delta N_{eff}}$ from different present and future 
generation experiments. The present $2\sigma$ limit from Planck(2018) on 
${\rm \Delta N_{eff}=0.285}$ excludes the DM mass between 
$\sim 1$ and $2$ MeV. Whereas the 1$\sigma$ limit 
from Planck(2018) on ${\rm \Delta N_{eff}=0.12}$ excludes DM mass in 
the range $\sim 3-4$ MeV.
It is also important to note that the future generation experiments such as 
the SPT-3G\cite{SPT-3G:2019sok} with 1$\sigma$ limit on $\Delta N_{\rm eff}=0.1$
and the CMB-S4\cite{CMB-S4:2016ple} with 2$\sigma$ limit on 
$\Delta N_{\rm eff}=0.06$ will be able to probe the 
DM mass upto $\sim 7$ MeV and $\sim 10$ MeV 
respectively. Hence with the 
present and future generation experimental measurement of 
$\Delta N_{\rm eff}$ we can indirectly probe the {\it freeze-in} DM 
in this scenario and also rule out certain mass range of such feebly
interacting dark matter. 
\section{Summary and Conclusion}
\label{sec:concl}
We have discussed a minimal extension of the SM by an inert $SU(2)_L$ scalar 
doublet $(\eta )$, a real scalar singlet $(S)$ and three right handed 
singlet fermions $(N_1,N_2, \& N_3) $ where all of them are odd under 
$\mathcal{Z}_2$ symmetry. However, the SM particles are even under 
the above-mentioned  $\mathcal{Z}_2$ symmetry. The study has been 
restricted to the regime where $M_{N_1}$ is lightest new particle 
in the spectrum and play the role of a stable DM candidate.
Due to the chosen $\mathcal{Z}_2$ symmetry, $N_i$s can only interact with SM 
particles through the yukawa interaction with $\eta$ and SM lepton doublet. 
We have assumed that the interaction of $N_1$ is very feeble which is 
decided by the Yukawa coupling $(y_{N_1}\lesssim 10^{-12})$. 
Such a small interaction of $N_1$ prevents 
its production in the thermal bath. Rather, ${N_1}$ has been produced from the
non-thermal decay of long lived  lightest scalar $\eta_1$, which is 
one of the neutral CP even scalar mass eigen state obtained by diagonalizing 
the $2\times2$ mass-square matrix in $(\eta^0, S)$ basis. The masses of all 
the other $\mathcal{Z}_2$ odd particles is sufficiently large and 
have no phenomenological consequences in the DM analysis. 
The SM neutrino masses can be generated through one loop process 
via interactions of $N_i$ and $\eta_1$ with the SM leptons. 
However, with such tiny Yukawa interaction ${N_1}$ is almost decoupled 
from the neutrino mass generation and as a result one of the 
active neutrino becomes almost massless. We have first checked the 
effects of 
different model parameters on the relic density of DM by solving the required 
Boltzmann equations. We have found that constraints coming from 
the electroweak precision data as well as the invisible 
decay of the SM Higgs boson ruled out a significant fraction of the model 
parameter space. We have seen that in the standard radiation 
dominated universe, the number densities of the mother particles~$(\eta_1)$ 
which have sizable interactions with the SM bath becomes 
smaller and have negligible 
impact on the $\Delta{N_{\rm eff}}$. On the other hand, if the early epoch 
of the expansion was dominated by some non-standard species that can 
chage the outcome by causing faster expansion of the universe. 
In this fast expanding universe scenario apart from producing the 
right amount of DM relic density, the late time 
decay of $\eta_1$ can significantly impact the total radiation energy 
density on the start of CMB formation and puts further constraints to 
the allowed parameter space of the model. In order to 
calculate the amount of ${\rm \Delta N_{eff}}$, we have found out the 
of the amount of energy density of neutrinos coming from the decay of 
$\eta_1$ by solving the required Boltzmann equation. The present $2\sigma$ 
limit on ${\rm \Delta N_{eff}}$ from Planck 2018 data excludes the DM mass 
as heavy as 2 MeV as in case of lighter DM mass, more and more energy gets 
converted to neutrinos and increases the value of ${\rm \Delta{N_{eff}}}$.
However, the future generation experiments like SPT-3G, CMB-IV will be
able to probe the dark matter mass as heavy as $\sim 10 $ MeV. Thus the analysis performed in this paper can be 
considered as an alternative way to probe the FIMP dark matter scenario
by the precise determination of $\Delta N_{\rm eff}$ using
the current and future CMB data.
\section{Acknowledgements}
 DN and SJ would like to thank D. Borah, A. Biswas, K. Dutta, S. Ganguly, D. Ghosh 
and Md. I. Ali for various discussions during the project. The work of DN 
is partly supported by National Research Foundation of Korea (NRF)’s grants, 
grants no. 2019R1A2C3005009(DN). The work of SJ is supported by CSIR, 
Government of India, under the NET JRF fellowship scheme
with Award file No. 09/080(1172)/2020-EMR-I.

\appendix

\section{Particle contents and the model parameters} {\label{apps}}
Particle content is given by
\begin{center}
\begin{tabular}{ |c|c|c|c| } 
 \hline
   & $SU(2)_L$ & $U(1)_Y$ & $Z_2$\\
   \hline
  $\phi$ & 2 & $\frac{1}{2}$& + \\ 
  \hline
 $\eta$ & 2 & $\frac{1}{2}$ & - \\ 
 \hline
 S &1 & 0 & - \\ 
 \hline
\end{tabular}
\end{center}
The scalar potential $V(\phi, \eta, S)$ is :
\begin{eqnarray} \nonumber
V(\phi, \eta, S) &=&-\mu_{\phi}^2 (\phi^{\dagger} \phi)+\mu_{\eta}^2 (\eta^{\dagger} \eta)+ \lambda_1 (\phi^{\dagger} \phi)^2+\lambda_2 (\eta^{\dagger} \eta)^2 + \mu_s^2 S^2+\lambda_s S^4\\ \nonumber &&+ \lambda_3 (\phi^{\dagger} \phi)(\eta^{\dagger}\eta )+ \lambda_4 (\phi^{\dagger} \eta)(\eta^{\dagger} \phi)+\frac{\lambda_5}{2} \left\{ (\phi^{\dagger} \eta)^2 + (\eta^{\dagger} \phi)^2\right\}\\ &&+\lambda_{\phi S} (\phi^{\dagger} \phi)S^2
+\lambda_{\eta S} (\eta^{\dagger} \eta)S^2 +\mu' \left\{(\phi^{\dagger} \eta)S+ (\eta^{\dagger} \phi)S \right\},
\label{potential}
\end{eqnarray}
where all parameters are real and $\mu_{i(i = \phi, \eta, s)}$ 
are the bare mass terms, $\mu^\prime $ is the trilinear scalar coupling, 
while various quartic scalar couplings are represented by 
$\lambda_{i(i =1-5)}, \lambda_{\phi S} $ and $\lambda_{\eta S}$ respectively. 
We write the SM scalar doublet as 
 \begin{equation}
 \phi=\dfrac{ 1 }{\sqrt{2}}
    \begin{bmatrix}
     & \phi_1 +i\, \phi_2 \\
          & \phi_3 + i\,\phi_4
    \end{bmatrix}\,
\end{equation}
where $\phi_i$ are real. From the minimization condition of the potential 
$V$ in equation \eqref{potential} we get:
\begin{equation}
\phi_j \left( -\mu_{\phi}^2 + \lambda_1 \sum_{i}^4 \phi_i^2 \right)=0    
\end{equation}
Any point on the circle $ \left( -\mu_{\phi}^2 + \lambda_1 \sum_{i}^4 \phi^2       \right)=0$ is a local minimum of the potential in equation \eqref{potential} and choosing a particular point ($\phi_1=\phi_2=\phi_4=0,\phi_3= v= \sqrt{\mu_{\phi}^2/ \lambda_1}$) will spontaneously break the symmetry.
 After the spontaneous symmetry breaking of the SM higgs doublet the doublet scalars can be represented as follows:\\
\begin{equation}
  \phi=
    \begin{bmatrix}
     & 0 \\
          &\dfrac{ v +h }{\sqrt{2}}
    \end{bmatrix}\,,
    \eta=
    \begin{bmatrix}
     & \eta^{\pm} \\
          &\dfrac{ \eta^0 + i\,A^0}{\sqrt{2}}
    \end{bmatrix},
\label{eq:h2d}
\end{equation}

where, $v = 246 $ GeV, is the SM electroweak vacuum expectation value (vev).
The masses of the SM like Higgs ($h$) and the charged scalar($\eta^{\pm}$) and the pseudo scalar particles($A^0$) can be written as ,
  \begin{eqnarray}
    M_{h}^2 & = &  2 \lambda_1 v^2 \\
    M^2_{\eta^{\pm}} & = & \mu_{\eta}^2 + \dfrac{\lambda_3}{2} v^2\\
    M^2_{A^0} & = &\mu_{\eta}^2 +(\lambda_3 + \lambda_4 -\lambda_5) \dfrac{v^2}{2}
\label{eq:etaA}
\end{eqnarray}
Due to the presence of the trilinear interaction $(\phi^{\dagger} \eta \, S)$ in the scalar potential, the neutral CP even component $\eta^0$ mixes with the real singlet scalar $S$ and the corresponding mass-square matrix in ($\eta^0$, $S$) basis has the following form:

\begin{equation}
M_{\eta^0 S}^2
=
 \begin{bmatrix}
 & \frac{\partial^2 V}{\partial {\eta^0} ^2} & \frac{\partial^2 V}{\partial {\eta^0} \partial S} \\
 &\frac{\partial^2 V}{\partial {\eta^0} \partial S} & \frac{\partial^2 V}{\partial S^2} 
 \end{bmatrix}
=
 \begin{bmatrix}
 & \mu_{\eta}^2+ \lambda_L v^2 & v\ \mu' \\
 & v\ \mu' &  2 \mu_s^2 + \lambda_{\phi S}\ v^2
 \end{bmatrix},
\label{eq:mass_mat}
\end{equation}
where, $ \lambda_L =\frac{1}{2}(\lambda_3 +\lambda_4+ \lambda_5)$.
The eigen values and CP-even neutral physical eigen states can be 
obtained by digonalising the above mass squared matrix: 
\begin{equation}
M^2_{\eta_1 \eta_2}=O^T M_{\eta^0 S}^2 O,
\label{eq:ortho}
\end{equation}
where 
\begin{equation}
O=
\begin{bmatrix}
\cos\theta & \sin{\theta}\\
 -\sin{\theta} & \cos{\theta}
\end{bmatrix},
\end{equation}
and $M^2_{\eta_1 \eta_2}$ is given by, 
\begin{equation}M^2_{\eta_1 \eta_2}=
\begin{bmatrix}
& M^2_{\eta_1} & 0\\
& 0 & M^2_{\eta_2}. 
\end{bmatrix} 
\label{eq:mass_mat1}
\end{equation}
The corresponding eigen states $\eta_1 $ and $\eta_2$ are given 
in terms of the weak basis $(\eta^0, S)$ and the mixing angle $\theta$:
\begin{eqnarray}
\eta_1 &=&  \eta^0 \cos\theta - S \sin\theta \\
\eta_2 &=&  \eta^0 \sin\theta + S \cos\theta 
\end{eqnarray}
The free parameters of the scalar sectors are the following:
$M_{\eta_1} $, $M_{\eta_2}$, $M_{\eta^\pm}$, $M_{A^0}$, $\lambda_\eta$, 
$\lambda_S$, $\lambda_{\phi S}$, $\lambda_{\eta S}$, $\lambda_3$, and 
$\sin \theta $. We set the SM like Higgs boson mass, $M_h = 125 $ GeV through 
out this analysis. All other parameters of the scalar sector can be
expressed in terms of those free parameters. 
From equations (\ref{eq:mass_mat}-\ref{eq:mass_mat1}) we get the following relations:
\begin{eqnarray}
\mu_{\eta}^2&=& \cos^2{\theta}\, M^2_{\eta_1}+  \sin^2{\theta}\, M^2_{\eta_2} -\lambda_L v^2\\
\mu'&=&\cos{\theta}\  \sin{\theta}\, \frac{1}{v} \left(M^2_{\eta_2}-M^2_{\eta_1}\right)\\
\mu_s^2&=& \frac{1}{2}(\sin^2{\theta}\, M^2_{\eta_1}+ \cos^2{\theta}\,  M^2_{\eta_2} - \lambda_{\phi S}\ v^2 )
\label{eq:relation}
\end{eqnarray}
Substituting these parameters in scalar masses in equation \eqref{eq:etaA} we get,
\begin{eqnarray}
 \lambda_5&=& \frac{1}{v^2}\left(\cos^2{\theta}\ M^2_{\eta_1} + \sin^2{\theta}\ M^2_{\eta_2} -M_{A^0}^2 \right) \\
 \lambda_4+\lambda_5 &=& \frac{2}{v^2}\left(\cos^2{\theta} M_{\eta_1}^2+\sin^2{\theta} M_{\eta_2}^2 -M_{\eta^+}^2 \right) 
 \label{params}
\end{eqnarray}
Now we have written the dependent parameters ($\mu_{\eta}\, ,\mu' \, ,\mu_S\, , \lambda_4\, , \lambda_5 $) in terms of free parameters.

\bibliographystyle{JHEP}
\bibliography{ref}

\end{document}